\documentclass[aps,pra,preprint,groupedaddress]{revtex4-1}

\usepackage{amsmath,amsfonts,amsthm} 
\usepackage{graphicx}
\usepackage{xcolor}

\begin{document}

\title{Skyrmion-string defects with arbitrary topological charges in spinor Bose-Einstein condensates}

\author{R. Zamora-Zamora and V. Romero-Roch\'in}
\email[]{romero@fisica.unam.mx}

\affiliation{Instituto de F\'isica, Universidad Nacional Aut\'onoma de M\'exico \\
Apartado Postal 20-364, 01000 Cd. M\'exico, Mexico}

\date{\today}

\begin{abstract}

Under the presence of external magnetic fields with cylindrical symmetry, Skyrmion-string defects with arbitrary topological charges are shown to appear in spinor $F = 1$ Bose-Einstein condensates. 
We show that, depending on the magnetic field boundary condition, the topological spin texture, at the planes perpendicular to the cylindrical axis, can take zero, half integer, or arbitrary values between $-1/2$ and $1/2$. We argue that these are true topological defects since their charge is independent of the spatial location of the singularity and since the total Skyrmion charge is the sum of the individual charges of the defects present. Our findings are obtained by numerically solving the corresponding fully coupled Gross-Pitaevskii equations without any symmetry assumptions. We analyze, both, polar $^{23}$Na and ferromagnetic $^{87}$Rb condensates.
\end{abstract}

\pacs{}

\maketitle

\section{Introduction}

The appearance of Skyrmions \cite{Skyrme127} range from systems in nuclear physics \cite{Brown20101}, superconductivity \cite{superconductivity}, magnetic solid state physics \cite{Nagaosa2013899,Mansoor2014}, liquid crystals \cite{liquidcrystals}, and Bose-Einstein condensates (BEC), thus becoming a topological unifying framework for seemingly different physical phenomena. In BEC systems, starting from early proposals to create and observe topological defects \cite{Leonhardt2000,PhysRevA.62.013602,PhysRevLett.88.090404,skyrmionOnHalfSpinor,PhysRevLett.88.080401} to recent ones \cite{PhysRevA.93.033633,PhysRevLett.109.015301,PhysRevLett.100.180403,Liu20133300,BorghPRL2016} carried on several experimental realizations already been achieved\cite{PhysRevLett.103.250401,PhysRevLett.108.035301,Hall2016,Ray2014657,LeanhardtPRL2003,ChoiNJP2012}. In particular, for the spinor $F = 1$ BEC that concerns us here, an intense recent activity on Skyrmions has developed \cite{XuPRA2012,PhysRevLett.108.035301,HuangPRA2013,BorghPRL2016}.

Skyrmions are topological defects of a given spatial vector field or order-parameter of the system in question, that can be classified in terms of homotopy groups of the corresponding field \cite{RevModPhys.51.591,topoBook,uedaAspectsSBEC}. There are versions in two- and three-dimensions and, typically, the Skyrmion topological charge is either an integer or a half-integer \cite{Leonhardt2000,mhVortex,vorticesTopo}. For a three-dimensional spinor $F = 1$ Bose-Einstein condensate, the field that develops the Skyrmion defects is the spin texture, a real measurable quantity.
We analyze different phases in the mentioned spinor BEC, that show a two-dimensional, and the same, Skyrmion topological defect in all planes perpendicular to a privileged axis; thus our naming of ``Skyrmion-string'' \cite{baby}. While this system has been the focus of great attention \cite{skyrmionOnHalfSpinor,XuPRA2012,HuangPRA2013,HuPRA2015,SuPRA2012,RuostekoskiPRL2001,ChoiNJP2012,BorghPRL2016}, we report here a novel aspect regarding the value of the acquired Skyrmion charge, namely, that depending particularly on boundary conditions of the external magnetic field that nucleates the vortices and Skyrmions in the spinor BEC, the 2D Skyrmion charge in each plane may take arbitrary values.
As we discuss in detail below, while the appropriate topological identification of the defect shown may need further elucidation in terms of topology theory, we support our claim that the defects are of a topological nature since their charge is independent of the location of the defect and, when there are more than one defect present, the total charge is the sum of the individual charges.
  
 Our study is based on the full numerical solution of the 3D spinor $F = 1$ Gross-Pitaevskii (GP) equations, without assuming any symmetry of the solution \cite{Zeng2009854}. We analyze both polar and ferromagnetic condensates, with parameters corresponding to actual values of $^{87}$Rb and $^{23}$Na \cite{RevModPhys.85.1191}. In addition to the numerical analysis, we discuss analytic results to support our discussion. Regarding the numerical study, we search for stationary states of $N$ confined, weakly interacting bosons of spin $F=1$, in the presence of an external magnetic field, whose energy functional is,
\begin{equation}
\begin{split}
E[\hat \Psi] &= \int d^3r \bigg[ \frac{\hbar^2}{2m}\nabla\Psi_n^*\cdot\nabla\Psi_n+V_{ext}(\vec{r})\Psi_n^* \Psi_n  \\&+\frac{c_0}{2}\Psi_n^* \Psi_n\Psi_k^* \Psi_k + \frac{c_2}{2}\Psi_k^* \vec F_{kn} \Psi_n \cdot \Psi_j^* \vec F_{jl} \Psi_l \\ &+p \vec{B} \cdot \Psi_n^* \vec{F}_{nk}\Psi_k \bigg] .\end{split}\label{totalEnergy} 
\end{equation}
The confining potential, $V_{ext} (\vec r)= m \omega^2 (x^2 + y^2 + z^2)/2$ is an isotropic harmonic optical trap with frequency $\omega = 2\pi \times 130 \>\text{Hz}$; $c_0$ and $c_2$ are the usual two-body interaction parameters as defined by Ho \cite{hoSpinor}, with $c_2 < 0$ for ferromagnetic phases and $c_2 > 0$ for polar ones, in the absence of the external field $\vec B$. The vector $\vec F$ are the $F = 1$ angular momentum matrices. The latin subindices run over the three components of spin $F = 1$. The last term is the linear Zeeman coupling to the external magnetic field $\vec B$, with strength $p < 0$. 
  
We consider external magnetic fields of the form, $\vec B = {\cal B}_0 \left( (x-x_0) {\bf x} - (y-y_0) {\bf y}\right) + B_z(r) {\bf z}$, with $r = ((x-x_0)^2 + (y-y_0)^2)^{1/2}$ and $(x_0,y_0)$ an arbitrary location inside the condensate planes. Two different types of the $z$-component are studied. In one case, $B_z(r) =$ constant, which can take any value, including zero. In the second case, $B_z(r) = {\cal B}_z r$, with ${\cal B}_z \ne 0$. The main difference of these two fields is their behavior as $r \to \infty$. In the first case, the direction of the field $\vec B/|\vec B| \to {\bf x} \cos \phi - {\bf y} \sin \phi$, with $\tan \phi = y/x$ the planar polar angle; that is, if $B_z = $ constant, the $\vec B$ field lies always on the $xy$-plane as $r \to \infty$. However, in the second case, the asymptotic direction of the $\vec B$ field no longer points on the plane, but in a direction defined by the values of ${\cal B}_0$ and ${\cal B}_z$,
\begin{equation}
\frac{\vec B}{|\vec B|} \to \frac{ {\cal B}_0 ({\bf x} \cos \phi - {\bf y} \sin \phi) + {\cal B}_z {\bf z}} {\left({\cal B}_0^2+{\cal B}_z^2\right)^{1/2}} \>\>\>\text{as}\>\>\>r \to \infty .\label{boundary}
\end{equation}
Our numerical analysis shows that in the presence of the $\vec B$ field, the polar and ferromagnetic character is lost as $r \to \infty$, the gas behaving as ``paramagnetic" with the spin texture pointing along the direction of the field. Thus, as we show below, the first case yields Skyrmions with charges 0, and $\pm 1/2$ always, while the second one gives rise to Skyrmions with arbitrary non-integer charge. In both cases, due to the zero of $\vec B$ along the line $(x_0,y_0)$, there appear well defined vortices with charges $0, \pm 1$, and $\pm 2$. \cite{nosostrosOnDemand} This is discussed in Section II. In Section III, we present the analysis concerning the Skyrmions topological charges, and we conclude with some remarks in Section IV.

\section{Vortices and Skyrmions in spinor condensates}

As already known \cite{LeanhardtPRL2003,PhysRevA.76.023610,nosostrosOnDemand,PhysRevA.61.063610}, magnetic fields of the type here considered, generate quantized vortices on the spin condensate components, along the line $(x_0,y_0)$. As it is shown in Ref. \cite{nosostrosOnDemand}, since the line or lines of zero field are at our disposal, one can create vortices, of the Mermin-Ho type \cite{mhVortex}, at arbitrary locations. Here, we show that a Skyrmion-string is also created at the same lines of zero field and, hence, they can also be externally created on demand. 

\begin{figure}[h!]
\centering
\begin{minipage}[b]{0.49\textwidth}
    \includegraphics[width=0.99\linewidth]{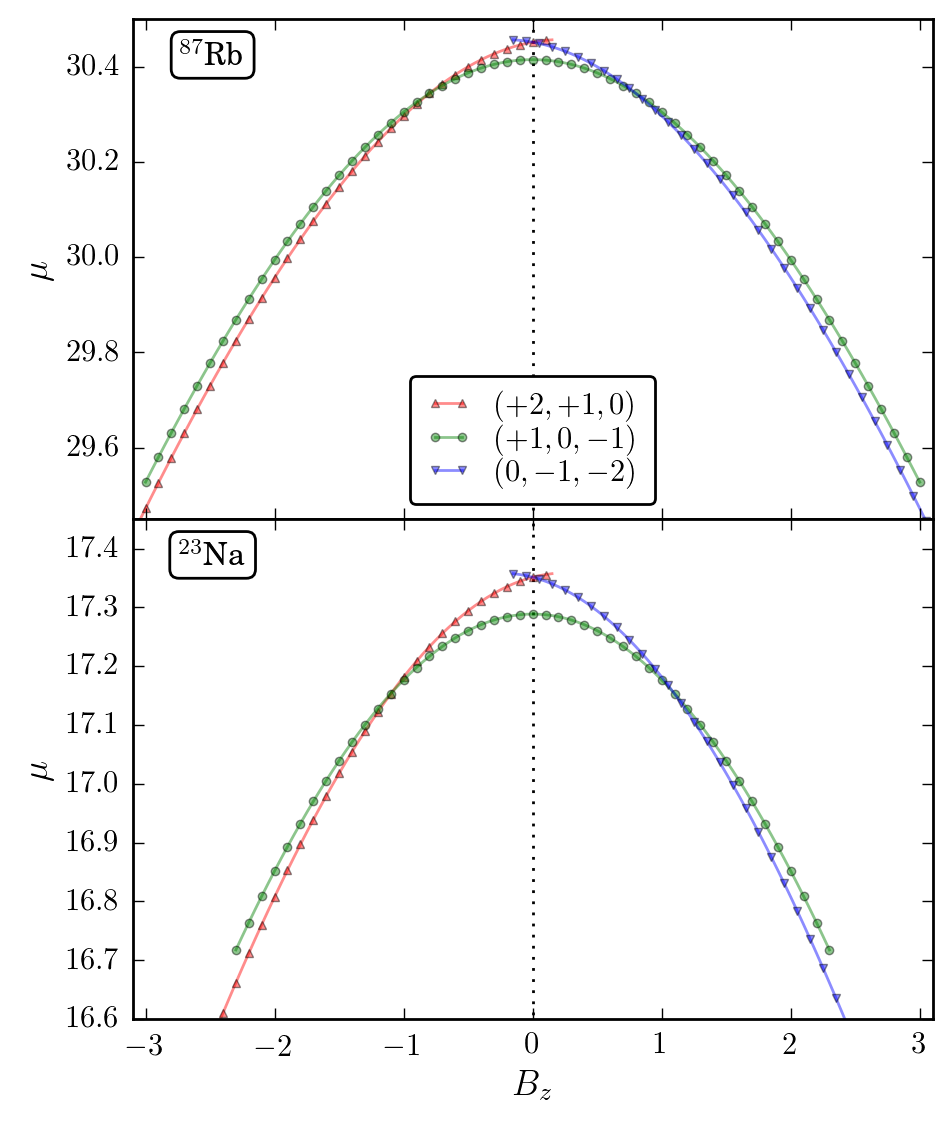}
  \end{minipage}
  \begin{minipage}[b]{0.49\textwidth}
   \includegraphics[width=0.99\linewidth]{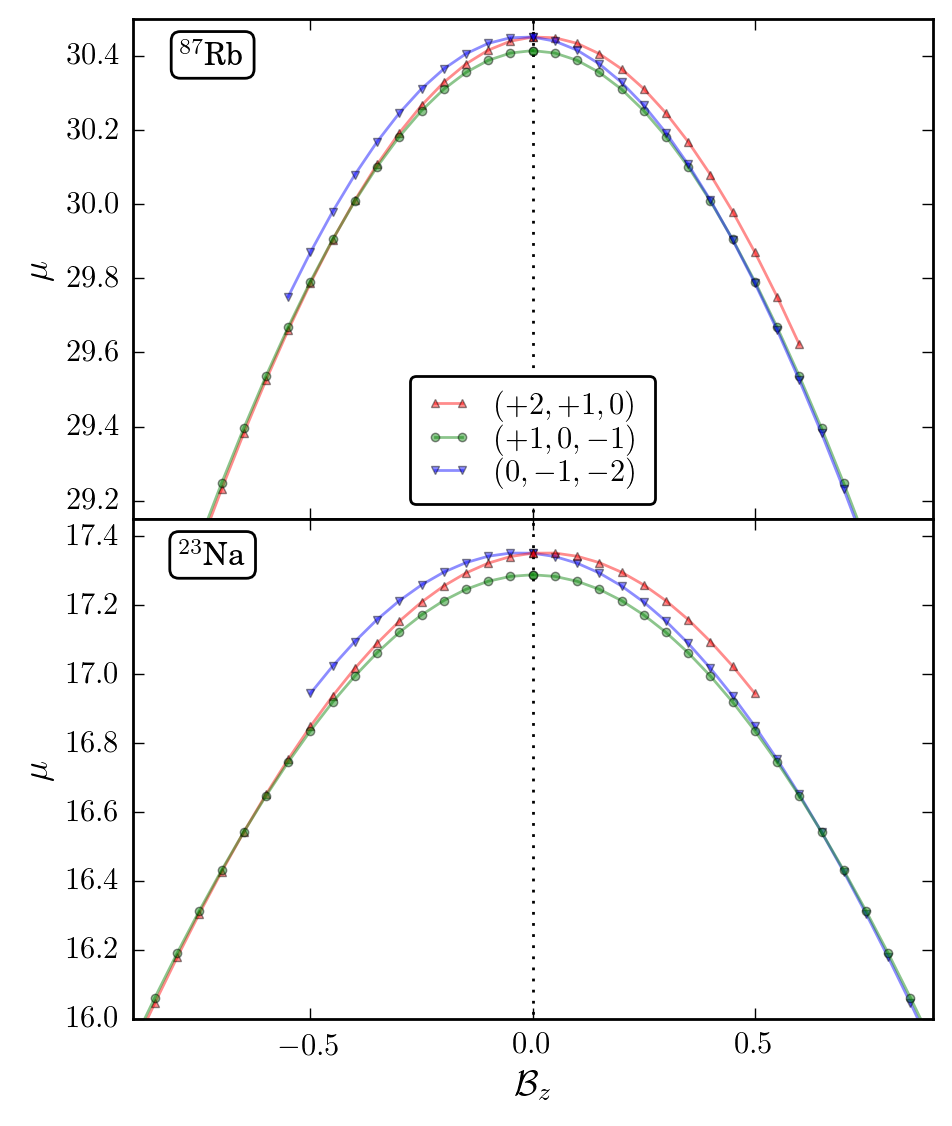}
  \end{minipage}
\caption{(Color online). Chemical potential $\mu$ as a function of $B_z =$ constant, left panel, and as a function of ${\cal B}_z$, right panel, for the case $B_z = {\cal B}_z r$, both for a ferromagnetic condensate $^{87}$Rb and a polar one $^{23}$Na. The labels (+2,+1,0), (+1,0,-1) and (0,-1,-2), indicate the three different vortex phases. These states are numerically stable, see text.}
\label{Sky05}
\end{figure}

Our numerical solutions of the corresponding GP equations indicate the existence of two or three stationary states \cite{calculation}, that may be called ``ground'' and ``excited'' states, depending on their corresponding chemical potential values. In order to classify those states we use the notation for Mermin-Ho vortices, namely, we find states $(1,0,-1)$, $(0,-1,-2)$ and $(+2,+1,0)$, whose notation $(l,m,n)$ represents vortices of charge $l$, $m$ and $n$ in the spinor components $m_F = +1, 0, -1$ of the field $\hat \Psi$. Fig. \ref{Sky05} shows the observed vortex-Skyrmion phases, in terms of the equilibrium chemical potential $\mu$ as a function of $z-$component of the external fields analyzed. The left panel of Fig. \ref{Sky05} refer to the case of $B_z =$ constant, while the right one to $B_z = {\cal B}_z r$, and in both cases we show the results for ferromagnetic ($^{87}$Rb) and polar ($^{23}$Na) phases. In both situations one sees that if $B_z = 0$, (+1,0,-1) is the ground state and $(+2,+1,0)$ and $(0,-1,-2)$ are degenerated excited states. As $B_z$ is turned on, the states (+1,0,-1) exist for both signs of $B_z$, while (+2,+1,0) exists always for $B_z < 0$, becoming (numerically) unstable for value of $B_z > 0$ above a threshold. The opposite is true for (0,-1,-2). We defer a discussion of the stability of the states to the last part of the article. 

\begin{figure}[h!]
\centering
\begin{minipage}[b]{0.3\textwidth}
    \includegraphics[width=1.1\linewidth]{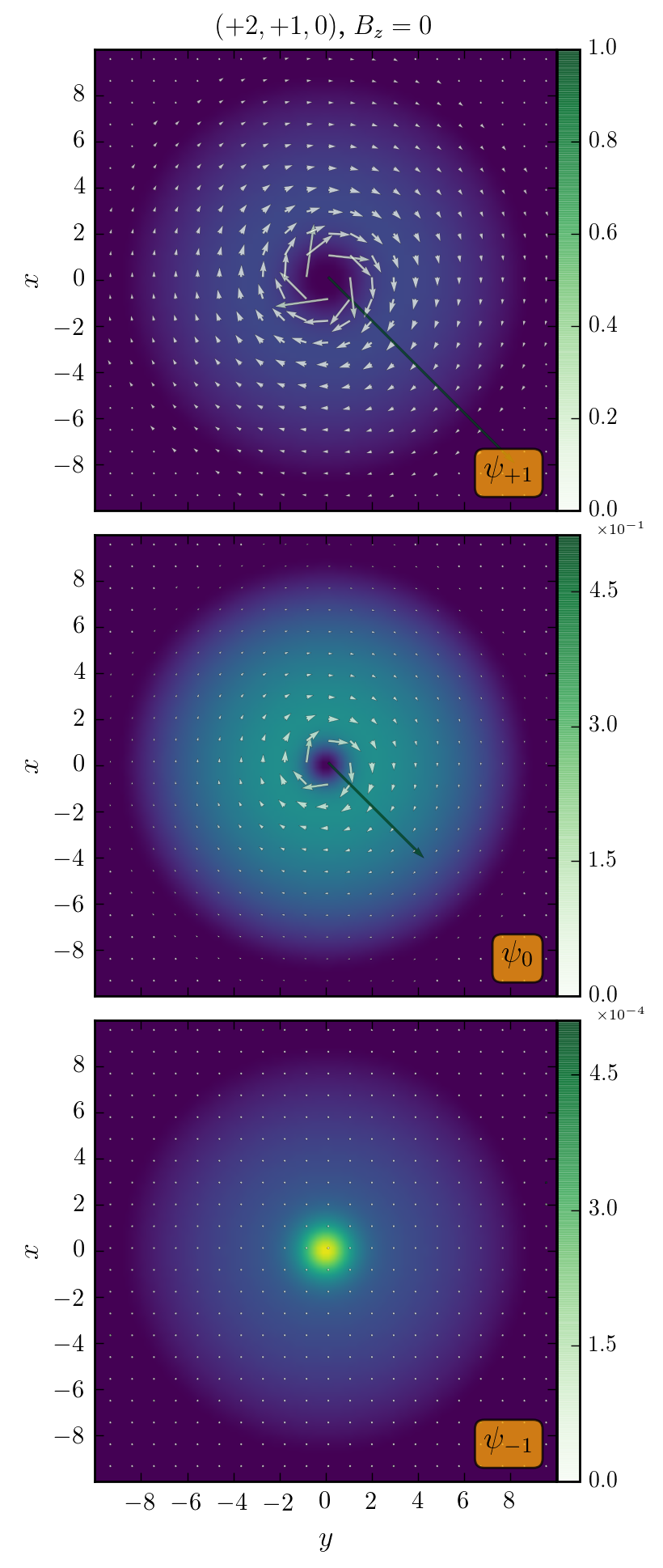}
  \end{minipage}
  \hfill
  \begin{minipage}[b]{0.3\textwidth}
   \includegraphics[width=1.1\linewidth]{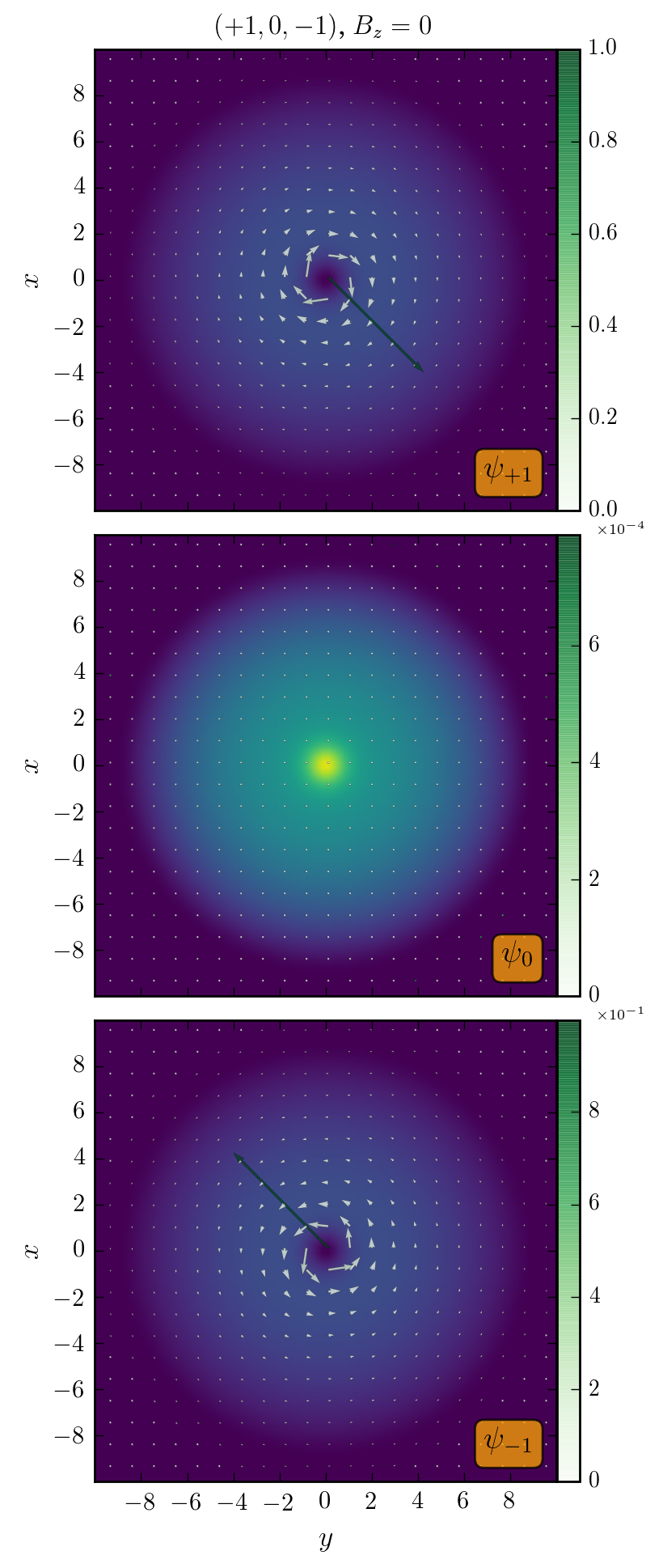}
  \end{minipage}
  \hfill
   \begin{minipage}[b]{0.3\textwidth}
   \includegraphics[width=1.1\linewidth]{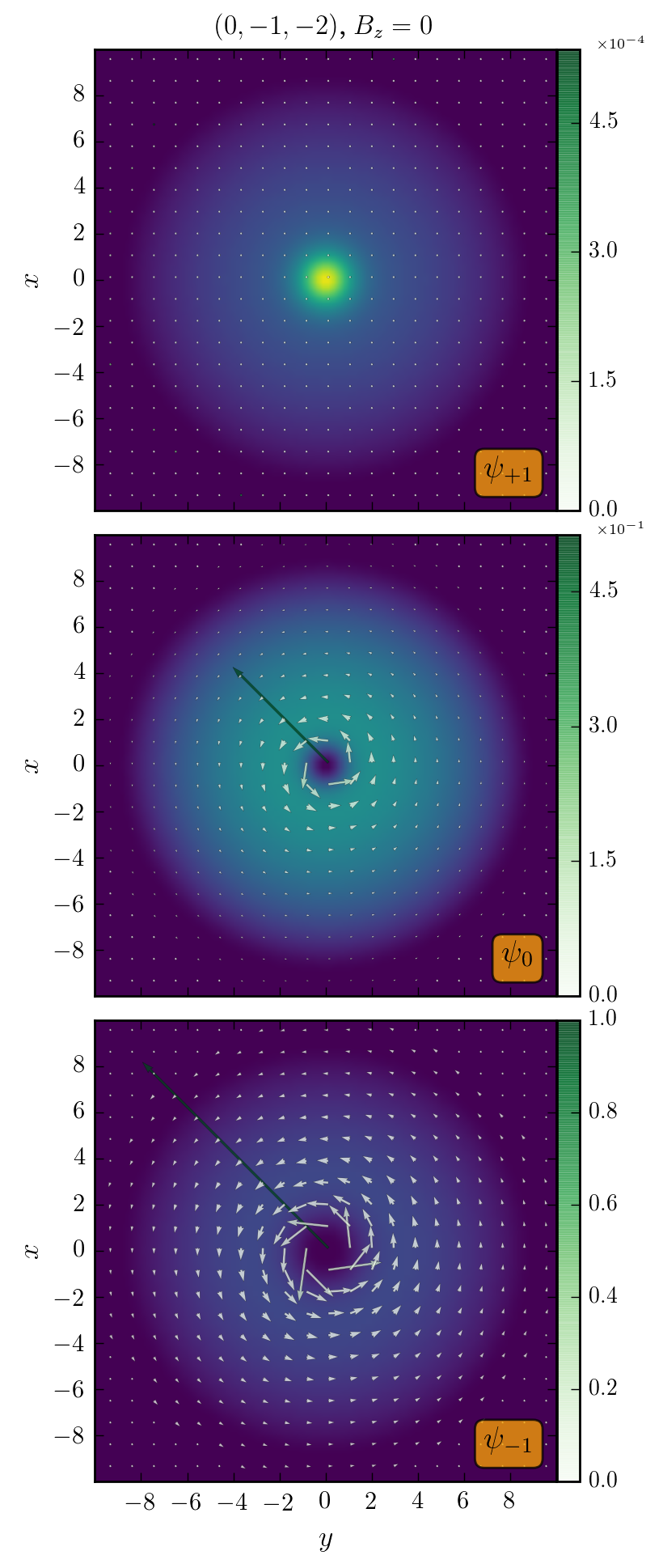}
  \end{minipage}
\caption{Typical velocity $\vec v_k$ and density fields $\rho_k$ of the condensate spinor components $\Psi_k$, in the $z=0$ plane. The density spike in the components with no vortices are due to the boundary condition of the spinor at $r = 0$, see Eqs. (\ref{101})-(\ref{012}).}
\label{Vortex-Bz-0}
\end{figure}

 The vortex classification of the above spinor BEC states can be fairly understood from the following considerations. The solution of the three GP equations for a $F = 1$ BEC can be written, in general, as
 \begin{equation}
\hat  \Psi(\vec r) = \sqrt{\rho(\vec r)} \left(
 \begin{array}{c}
 \zeta_{+1}(\vec r) e^{i \theta_{+}(\vec r)} \\  \zeta_{0}(\vec r) e^{i \theta_{0}(\vec r)}  \\  \zeta_{-1}(\vec r) e^{i \theta_{-}(\vec r)} 
 \end{array} \right) \label{PSI}
 \end{equation}
 where $\rho(\vec r) = \hat \Psi^\dagger(\vec r) \hat \Psi(\vec r)$ is the total particle density of the condensate. The amplitudes $\zeta_m(\vec r)$ are real functions, obeying $\zeta_{+1}^2(\vec r) + \zeta_{0}^2(\vec r) +\zeta_{-1}^2(\vec r) = 1$ everywhere.  An analysis of the vortex solutions, if they exist and if single valuedness is imposed, leads to the following general results \cite{nosostrosOnDemand}: 
(1) Only two components can show a vortex, say $\zeta_\alpha \to 0$ and $\zeta_\beta \to 0$ as $r \to 0$, with $\theta_\alpha \ne 0$ and $\theta_\beta \ne 0$, while the third one, $\zeta_\gamma \to 1$ as $r \to 0$, and $\theta_\gamma = 0$. And, (2) the phases differences obey $\theta_m - \theta_{m-1} = \phi$, with $\phi$ the polar angle. These two conditions imply the appearance of three vortex solutions, with boundary conditions at $r \to 0$ and with charges $\theta_m = Q_m \phi$, given by
\begin{eqnarray}
& (1,0,-1)& \nonumber \\
& \zeta_{+1} \to 0 \>,\> \zeta_{0} \to 1 \>,\> \zeta_{-1} \to 0 \>\>{\rm as} \>\>r \to 0  & \nonumber \\
& Q_{+1} = +1 \>,\> Q_{0} = 0 \>,\> Q_{-1} = -1  &. \label{101}
\end{eqnarray}
\begin{eqnarray}
& (+2,+1,0) & \nonumber \\
& \zeta_{+1} \to 0 \>,\> \zeta_{0} \to 0 \>,\> \zeta_{-1} \to 1 \>\>{\rm as} \>\>r \to 0  & \nonumber \\
&Q_{+1} = +2 \>,\> Q_{0} = 1 \>,\> Q_{-1} = 0  &. \label{210}
\end{eqnarray}
\begin{eqnarray}
& (0,-1,-2) & \nonumber \\
& \zeta_{+1} \to 1 \>,\> \zeta_{0} \to 0 \>,\> \zeta_{-1} \to 0 \>\>{\rm as} \>\>r \to 0  & \nonumber \\
& Q_{+1} = 0 \>,\> Q_{0} = -1 \>,\> Q_{-1} = -2  &.\label{012}
\end{eqnarray}
Fig. \ref{Vortex-Bz-0} shows the velocity field $\vec v_k = \frac{\hbar}{m} \nabla \Phi_k$ of the condensate spinor components $\Psi_k$, with $\tan \Phi_k = {\rm Im} \Psi_k/{\rm Re} \Psi_k$, as well as the their particle density $\rho_k = (\Psi_k^* \Psi_k)^{1/2}$, in the $z=0$ plane. The vortices and the corresponding density spikes, $\zeta_m \to 1$ as $r \to 0$, can be clearly observed.
For the other cases, $B_z = $constant and $B_z = {\cal B}_z r$, these vortex structures remain and, at first sight, look as in Fig. \ref{Vortex-Bz-0}. Certainly, as shown below, one can tailor external fields with more than one vortex per component.

We now turn to the spin texture description. For this, it first appears convenient to factorize the phase of the 0-component of the full solution $\hat  \Psi(\vec r)$, Eq. (\ref{PSI}), and write
 \begin{equation}
\hat  \Psi(\vec r) = \sqrt{\rho(\vec r)} e^{i \theta_0(\vec r)} \hat \zeta(\vec r) \label{choice}
 \end{equation}
 where
 \begin{equation}
 \hat \zeta(\vec r) = \left(
 \begin{array}{c}
  \zeta_{+1}(\vec r) e^{i \delta_{+}(\vec r)} \\  \zeta_{0}(\vec r)  \\  \zeta_{-1}(\vec r) e^{i \delta_{-}(\vec r)} 
 \end{array} \right)
 \end{equation}
with $\delta_m = \theta_m - \theta_0$. We shall leave the $\vec r$ dependence implicit in the foregoing analysis. As it is common,  a 3D complex vector $\vec a = (a_x,a_y,a_z)$ can be introduced, with $a_x=\frac{1}{\sqrt{2}}(\zeta_{-1}e^{i \delta_{+}} -\zeta_{+1}e^{i \delta_{-}})$, $a_y=\frac{-i}{\sqrt{2}}(\zeta_{-1}e^{i \delta_{+}} +\zeta_{+1}e^{i \delta_{-}})$, and $a_z=\zeta_0$, obeying $\vec{a} \cdot \vec{a}^*=1$. This vector can be further decomposed in its real and imaginary parts, $\vec a = \vec a_R + i \> \vec a_I$. 

The magnetization, a physical observable, is given by $\hat \Psi^\dagger \vec F \hat \Psi = \rho \vec f$, with $\vec f = \hat \zeta^\dagger \vec F \hat \zeta$ the spin texture. Using the above decomposition of the state, the spin texture can be expressed as $\vec f = 2 \> \vec a_R \times \vec a_I$. Let us analyze this expression for different cases. The simplest one is when the full Zeeman contribution is zero, namely $\vec B = 0$ in Eq. (\ref{totalEnergy}). The solution, as shown by Ho \cite{hoSpinor}, is that for the polar case, $c_2 < 0$, $\vec a$ is real and $\vec f = 0$. For the ferromagnetic case, $c_2 > 0$, $\vec f = \vec f_0$, a constant vector everywhere, which by an appropriate rotation can be brought to the case $\zeta_{+1} = 1$, $\zeta_0 = \zeta_{-1} = 0$, namely $\vec f = {\bf z}$. For $\vec B \ne 0$, the polar and ferromagnetic characters are overridden and the three types of quantum-vortex phases appear. 

The above spinor vortex structures have an additional associated topological defect that one may call a ``string-Skyrmion" due to the presence of the external magnetic field. That is, while the trap imposes its spherical symmetry on the total density $\rho(\vec r)$, the external magnetic field imposes its additional cylindrical symmetry on the texture field $\vec f$, as one should expect on physical grounds. Hence, the spin texture does not depend on the $z$ coordinate, $\vec f = \vec f(x,y)$, and furthermore, it shows cylindrical symmetry around the location of the zero line of the $\vec B$ field. Thus the ``string" qualifier. In other words, the spin texture shows the same geometric and topological structure in all $z-$planes .

The connection between the vortex solutions and the Skyrmions can be found as follows. An important step lies in the restriction on the phases of the vortex structure, $\theta_m - \theta_{m-1} = \phi$. This restriction yields the phase requirement $\theta_{+1} + \theta_{-1} - 2 \theta_0 = 0$, which in turn, and only in this case, makes the set $(\vec a_R, \vec a_I, \vec f)$ an orthogonal triad (group O(3)). Their explicit form is
\begin{equation}
\vec{a}_R = \frac{\zeta_{-1}-\zeta_{+1}}{\sqrt{2}} {\boldsymbol \rho} + \zeta_0 {\bf z}
\end{equation}
\begin{equation}
\vec{a}_I =  -\frac{\zeta_{-1}+\zeta_{+1}}{\sqrt{2}} {\boldsymbol \delta}
\end{equation}
and, hence,
\begin{equation}
\vec{f}=( \zeta_{+1}^2-\zeta_{-1}^2) {\bf z} + \sqrt{2}\zeta_0(\zeta_{+1}+\zeta_{-1}) {\boldsymbol \rho} .\label{ff}
\end{equation}
The unit vectors are ${\boldsymbol \rho} = {\bf x} \cos \delta + {\bf y} \sin \delta$ and ${\boldsymbol \delta} = - {\bf x} \sin \delta + {\bf y} \cos \delta$, with $\delta = \delta_- = - \delta_+$. The angle $\delta$ spans $0 \to 2 \pi$, since it is actually the phase of the wavefunctions that yield the vortex charges, however, it is not the polar angle $\tan \phi = y/x$. Nevertheless, the triad unit vector $({\boldsymbol \rho},{\boldsymbol \delta},{\bf z})$ together with $r = (x^2 + y^2)^{1/2}$ and $z$ span the space with cylindrical symmetry. This is important, as we shall return below, since far from the singularity, ${\boldsymbol \rho}$ tends to align to the in-plane component of the $\vec B$-field, namely, to $({\bf x} \cos \phi - {\bf y} \sin \phi)$. It is also important to mention that while $\vec a_R$ and $\vec a_I$ depend on the choice of the global phase, as shown in Eq. (\ref{choice}), $\vec f$ does not. This is because $\vec f$ is an observable and  $\vec a_R$ and $\vec a_I$ are not.

Since the (real) spinor components depend only on $r$, $\zeta_m = \zeta_m(r)$, the spin texture can be written as $\vec f = f_z(r) {\bf z} + f_r(r) {\boldsymbol \rho}$, and the components $f_z$ and $f_r$ can be read off of Eq. (\ref{ff}). With these, the existence or not of an Skyrmion structure in any plane $z = $constant, can be checked. To this end one recalls that the 2D Skyrmion charge is given by
\begin{equation}
Q_{sky}^{2D} = \frac{1}{4\pi} \iint dx dy \> \vec{f}\cdot\left(\frac{\partial \vec{ f}}{\partial x} \times \frac{\partial \vec{ f}}{\partial y}\right),
\label{magSky}
\end{equation}
which, with the cylindrical symmetry of $\vec f$, can be cast as,
\begin{equation}
Q_{sky}^{2D} = \frac{1}{2} \int_0^\infty \> f_r \left( f_z \frac{d f_r}{dr} - f_r \frac{d f_z}{dr} \right) dr . \label{sky}
\end{equation}
There exists, however, an additional important constraint for the cases studied in this work, except for the polar vortex (+1,0,-1), 
namely, that $\vec f \cdot \vec f = 1$ everywhere. Although we have not been able to prove this constraint rigorously, our numerical solutions demonstrate it. This restriction may be written as $f_z^2 + f_r^2 = 1$, which further implies that the Skyrmion charge, Eq.(\ref{sky}), can be cast as, 
\begin{equation}
Q_{sky}^{2D} = \frac{1}{2}  \left(f_z(0) - f_z(\infty) \right) \>\>\>{\rm if}\>\>\>  \vec f \cdot \vec f = 1. \label{sky-res}
\end{equation}
That is, if $\vec f \cdot \vec f = 1$ everywhere, the Skyrmion charge is given solely by (one-half) the difference of the boundary values of the $z-$component of the spin texture. The boundary value at $r = 0$ can always be found, as can be seen from Eqs. (\ref{101})-(\ref{012}) and (\ref{ff}), however, the boundary value at $r \to \infty$ cannot be directly accessed since the confining harmonic trap allows for calculations up to the Thomas-Fermi radius of the atomic cloud only. Below, along the presentation of our results, we show how we circumvent this difficulty and how we find trustable values of the topological charges.

\section{Skyrmions in magnetic fields with different boundary conditions}

\begin{figure}[h!]
 \centering
  \includegraphics[width=0.7\linewidth]{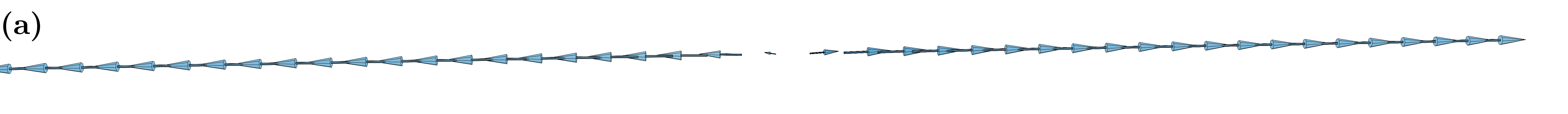}
 \includegraphics[width=0.7\linewidth]{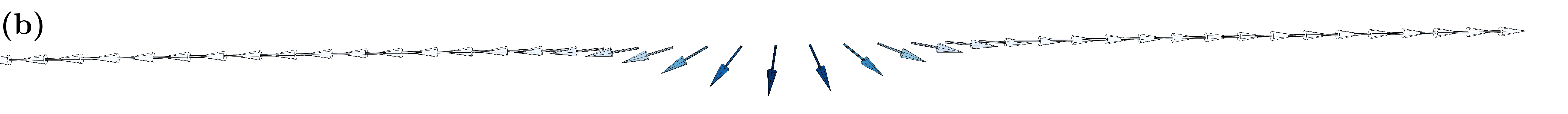}
  \includegraphics[width=0.7\linewidth]{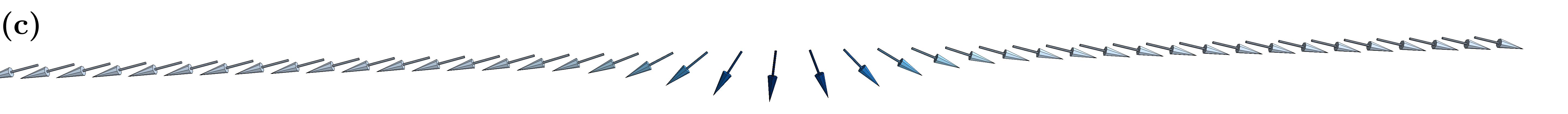}
 \includegraphics[width=0.7\linewidth]{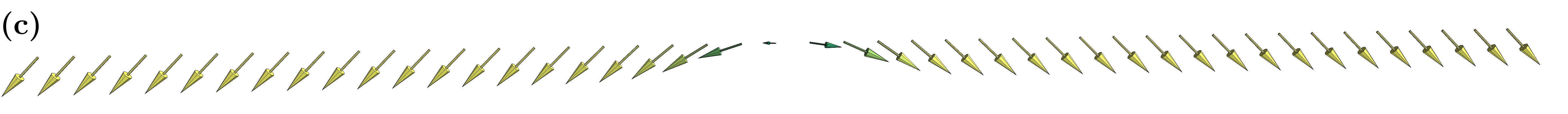}
  \includegraphics[width=0.7\linewidth]{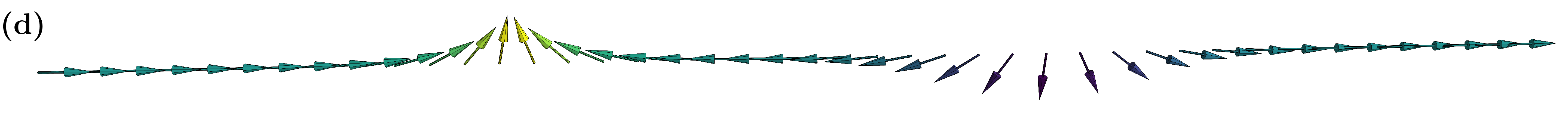}
  \includegraphics[width=0.7\linewidth]{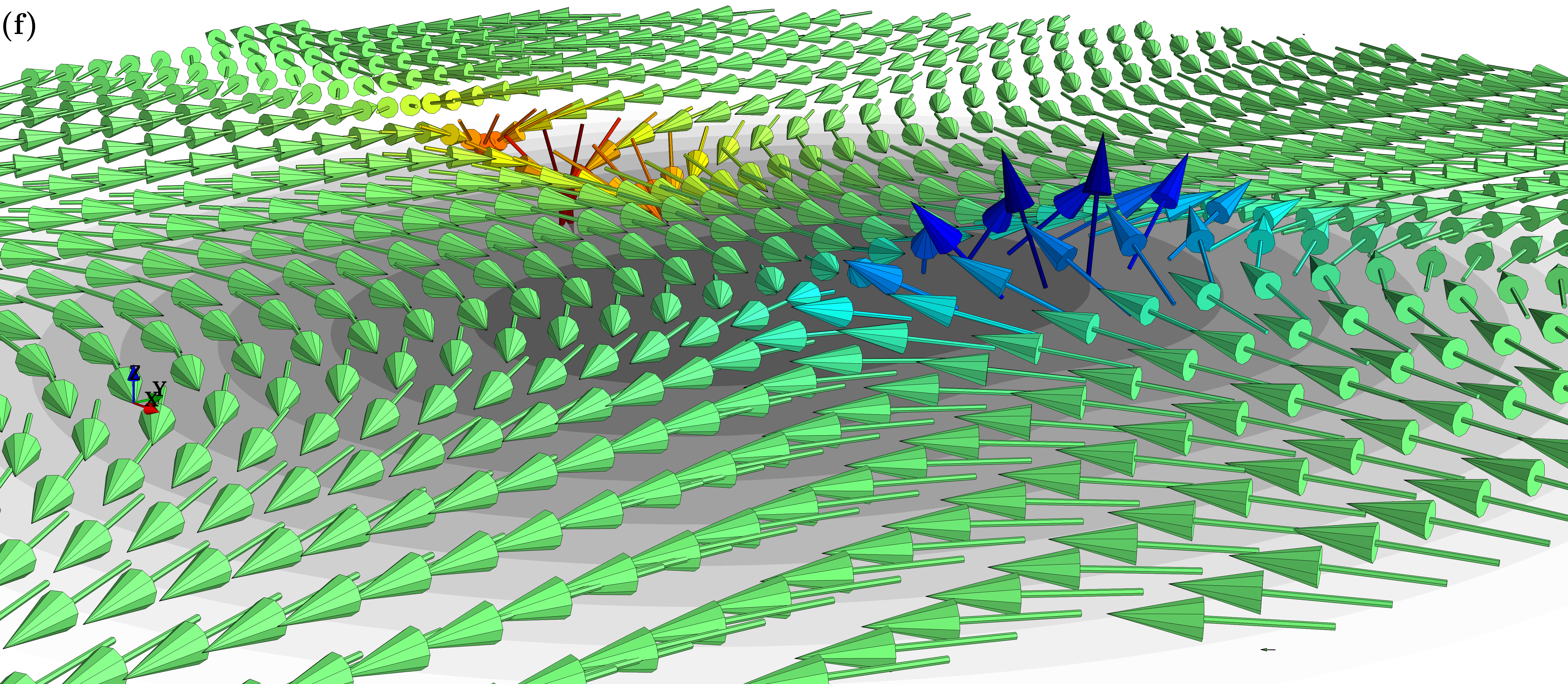}
\caption{Skyrmion spin textures $\vec f$. (a) (+1,0,-1), $Q_{sky}^{2D} = 0$, $B_z = 0$. (b) (+2,+1,0), $Q_{sky}^{2D} = -1/2$, $B_z = 0$. (c) (+2,+1,0), $Q_{sky}^{2D} = 0.37$ $B_z =-0.25 r$. (d) (+1,0,-1), $Q_{sky}^{2D} = - 0.19$. $B_z = -r$. (e) Two Skyrmions (+2,+1,0) + (0,-1,-2), $Q_{sky}^{2D} = +1/2 - 1/2 = 0$, $B_z = 0$. (f) 2D view of two Skyrmions, $B_z = 0$, the alignment of $\vec f$ to the $\vec B$-field far from the singularities.}  \label{penachos}
\end{figure}

As discussed in the previous section, the vortex structures (+2,+1,0), (+1,0,-1) and (0,-1,-2) each have associated their own spin textures and Skyrmion defects. Moreover, although the spin texture of the vortex (+1,0,-1) does not satisfy $\vec f \cdot \vec f = 1$ everywhere, because $\vec f \to 0$ as $r \to 0$, it is still clear that the boundary conditions are essential to determine the Skyrmion charges. The boundary conditions at $r = 0$ are completely determined by the corresponding vortex structure, as shown by Eqs. (\ref{101})-(\ref{012}). However, supported by our calculations, the boundary values at $r \to \infty$ depend on the external $\vec B$-field. We find that far from the vortex singularity,  $\vec f$ tends to align to the direction of the magnetic field $\vec B$. This can be further understood by looking at the expression for the energy $E[\Psi]$, Eq. (\ref{totalEnergy}), in which the Zeeman term can be written as $p \vec{B} \cdot \Psi_n^* \vec{F}_{nk}\Psi_k  = p |\vec B| \rho (\vec B/|\vec B|) \cdot \vec f$. Since $p < 0$ in our calculations, this term tends to minimize the energy when $\vec B/|\vec B|$ and $\vec f$ are parallel. We discuss separately the three general cases. Fig. \ref{penachos} illustrate the spin texture $\vec f$ for several typical Skyrmions.

\subsection{$B_z = 0$}

First, for completeness, we review the case in the absence of a $z$-component of the $\vec B$-field, namely, $B_z = 0$. For the vortex solution (+1,0,-1), by symmetry (and numerically verified) $\zeta_{+1}^2 = \zeta_{-1}^2$ for all $r$, hence, $f_z = 0$, see Eq.(\ref{ff}). That is, the vector $\vec f$ not only lies on the $xy$-plane, it becomes zero as $r \to 0$. Its Skyrmion charge is thus zero, $Q_{sky}^{2D}(+1,0,-1) = 0$. This is the so-called polar coreless vortex \cite{PhysRevA.76.023610}. 
For the vortices (+2,+1,0) and (0,-1,-2), Eqs. (\ref{210}), (\ref{012}) and (\ref{ff}), show that $f_r \to -1$ and $f_r \to +1$, respectively, as $r \to 0$. The numerical solutions further show that $\zeta_{+1}^2 = \zeta_{-1}^2$ as $r \to \infty$, namely $f_r \to 0$ as $r \to \infty$ for both cases. See Fig (+2,+1,0) as example of this case. Hence, one finds that the vortex (+2,+1,0) has an associated Skyrmion charge $Q_{sky}^{2D}(+2,+1,0) = -1/2$ and, analogously, the vortex (0,-1,-2) has $Q_{sky}^{2D}(0,-1,-2) = +1/2$. We also calculated these charge values by integrating directly $Q_{sky}^{2D}$, using the full equation (\ref{magSky}), finding values very close to $\pm 1/2$. 

\subsection{$B_z = {\rm constant}$}

\begin{figure}[h!]
 \centering
 \includegraphics[width=0.7\linewidth]{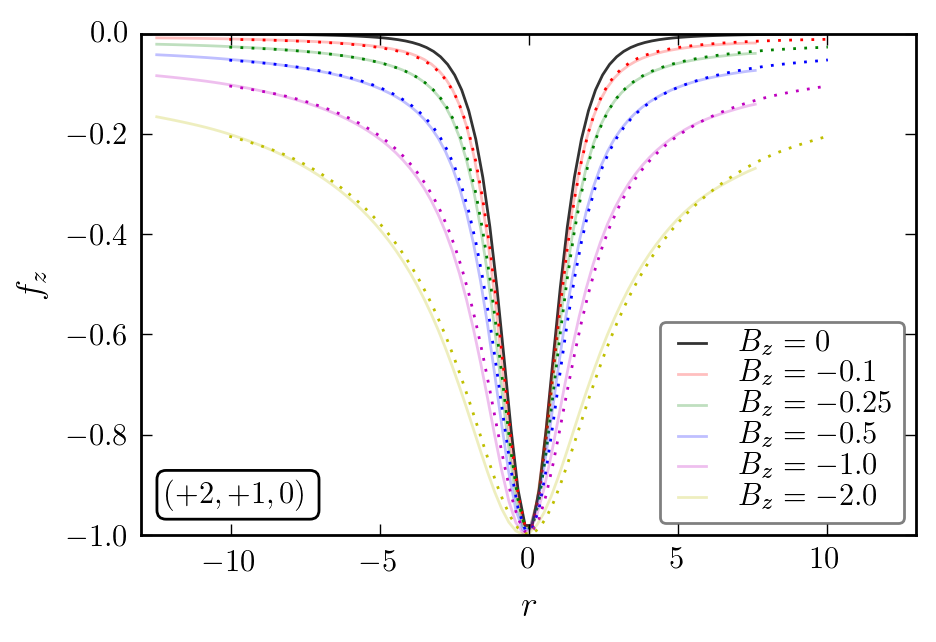} 
 \includegraphics[width=0.7\linewidth]{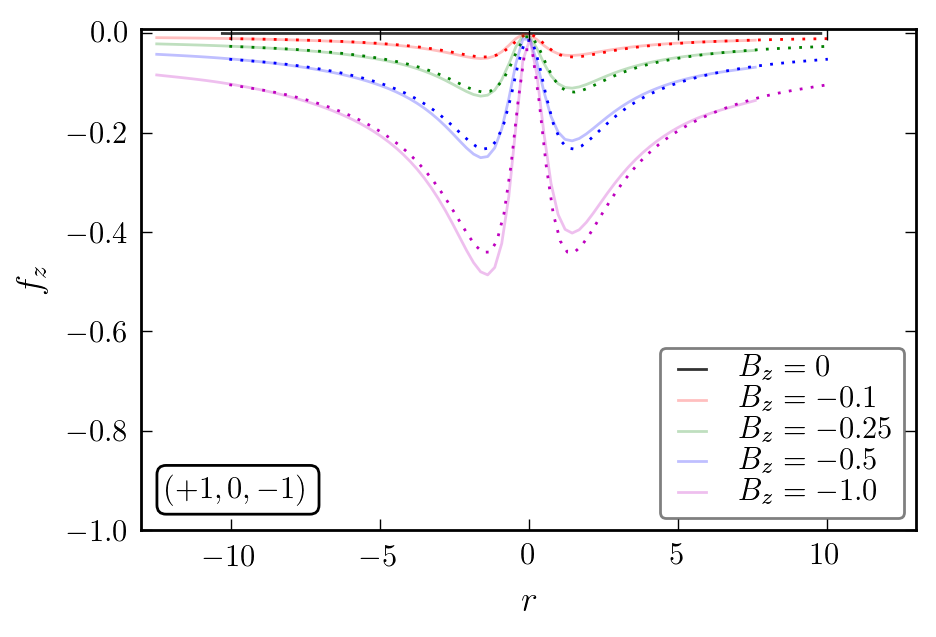}
 \caption{Spin texture component $f_z$ as a function of $r$, for different values of $B_z =$ constant, for the vortex structure (+2,+1,0), upper panel, and (+1,0,-1), lower panel. The  dotted lines corresponds to a magnetic field centered at $(x_0 =0,y_0 = 0)$, while the continuous ones to a $(x_0 = 0,y_0 =2)$. This constitutes an indication that $f_z \to 0$ as $r \to \infty$.}  \label{SKY210}
\end{figure}

\begin{figure}[h!]
 \centering
 \includegraphics[width=0.7\linewidth]{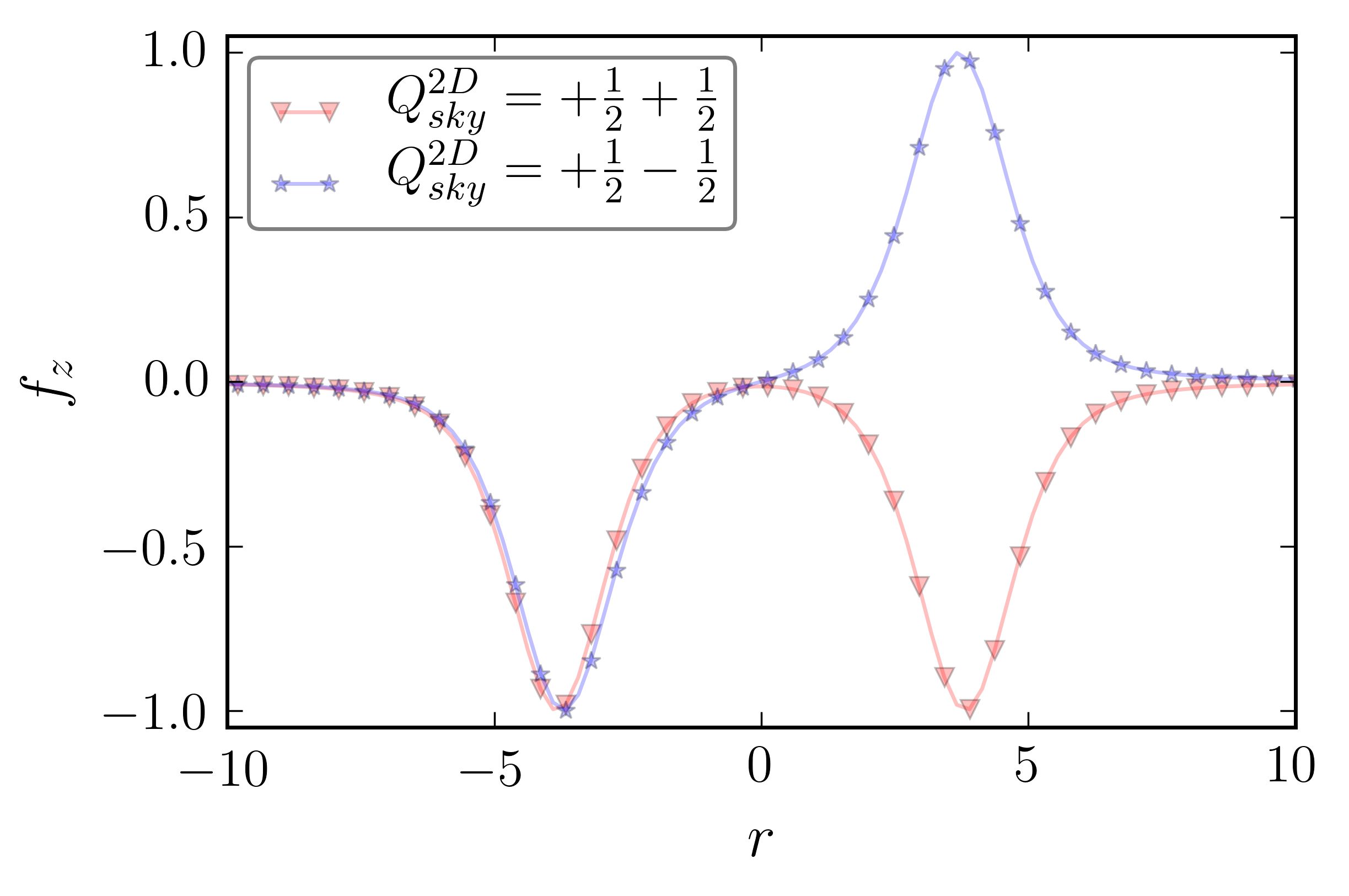}
\caption{Spin texture components $f_z$ as a function of $r$, for two defects, one located $(x_0 \approx -3.6,y_0 = 0)$ and the other at  $(x_0 \approx 3.6,y_0 = 0)$, with charges (+1/2, +1/2) and (+1/2-1/2), yielding $Q_{sky}^{2D} = 1$ and $Q_{sky}^{2D} = 0$, respectively.}  \label{DOS-SKY-BZ-CONST}
\end{figure}

We now consider a constant $z-$component of the external $\vec B$-field, but different from zero. If $B_z > 0$, the vortex structure remains the same as before but, as seen in Fig. 1, the most stable case is now (+2,+1,0). For $B_z < 0$, the situation is reversed and the stable phase is (0,-1,-2). A direct calculation of $Q_{sky}^{2D}$ using the definition given by Eq. (\ref{magSky}) shows that the charge is {\it apparently} not zero for the polar Skyrmion (+1,0,-1) and different from $\pm 1/2$ for (+2,+1,0) and (0,-1,-2) (Results not shown here). This is a numerical artifact, however, because the cloud reaches out up to the Thomas-Fermi radius only, namely, the particle density is numerically negligible beyond it. That is, notwithstanding that the structure of the spin texture $\vec f$ is modified by the $B_z \ne 0$ component, we assert that the topological charges remain 0 for the polar Skyrmion (+1,0,-1) and $\pm 1/2$ for (+2,+1,0) and (0,-1,-2). To verify this, we calculated the spin texture for the cases where the $\vec B-$field has its zero line at $(x_0=0,y_0=0)$ and also at $(x_0\ne 0,y_0\ne 0)$. As expected, the vortex structure remains the same, except that the vortices are centered now at the corresponding values $(x_0,y_0)$. To see the effect on the Skyrmions we show Fig. \ref{SKY210}. The upper panel shows the $z-$component of the spin texture, $f_z(r)$, for vortices (+2,+1,0) with different values of $B_z$, and in which we have superposed the solutions with $(x_0=0,y_0=0)$, marked with dots, with those of $(x_0\ne 0,y_0\ne 0)$, with continuous lines. It can be clearly seen that as $r$ grows, $f_z$ keeps diminishing with no bound. We take this as an evidence that as $r \to \infty$, it must be true that $f_z \to 0$ and, therefore, that the topological Skyrmion charge is $-1/2$ in this case. Incidentally, the fact that the charge is the same, independently of the location of the defect, tell us that it is of a topological nature. For the case (+1,0,-1), lower panel in Fig. \ref{SKY210}, we see that the spin texture structure changes strongly from $B_z = 0$ to $B_z \ne 0$, however, the topological charge remains zero. We conclude this from a direct calculation of the Skyrmion charge for larger values of the cutoff of the radial integral Eq.(\ref{sky}). 
The explanation that as $r \to \infty$, $f_z(r) \to 0$, for all these cases, thus yielding charges 0 or $\pm 1/2$ independently of the value of $B_z =$constant, is that far from the defect location, $(x_0,y_0)$, the spin texture $\vec f$ becomes paramagnetic in the sense that it gets aligned to the direction of the external $\vec B$-field. Since this field points along the $xy$-plane as $r \to \infty$, hence $f_z(r) \to 0$. This insight suggests to consider the other case of the $\vec B$-field, which can be made to point at arbitrary directions, such that the boundary value of $f_z(r)$, at $r \to \infty$, can be also made to point to such a direction. For completeness and to reinforce the topological nature of the defects, we plot in Fig. \ref{DOS-SKY-BZ-CONST} the profile of two Skyrmion defects of the same type in the cloud, one case with the same sign of the charges and the other with opposite signs, thus yielding the double of the charge and zero respectively. 

\subsection{$B_z = {\cal B}_z r$}

\begin{figure}[h!]
 \centering
 \includegraphics[width=0.7\linewidth]{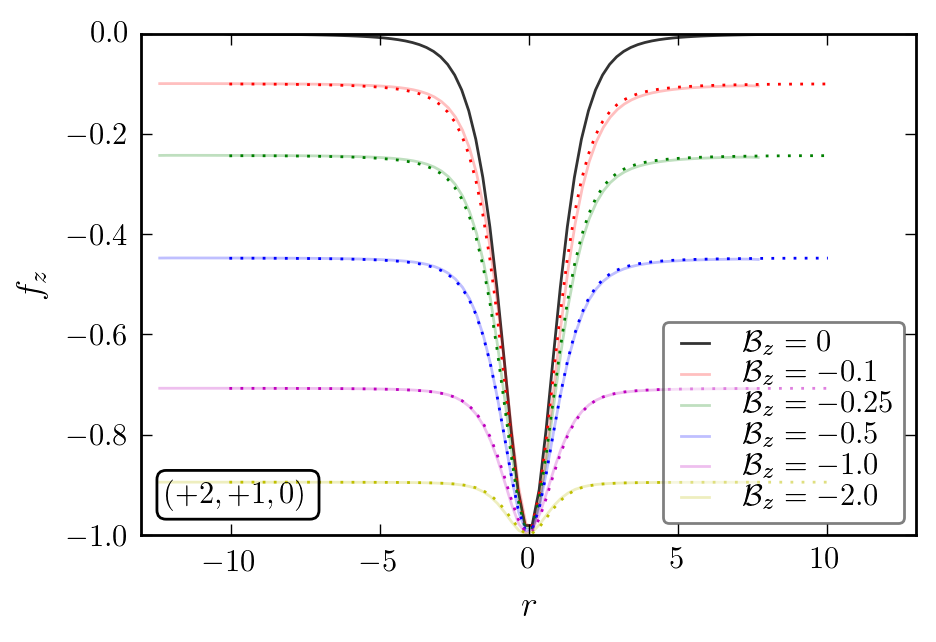}
  \includegraphics[width=0.7\linewidth]{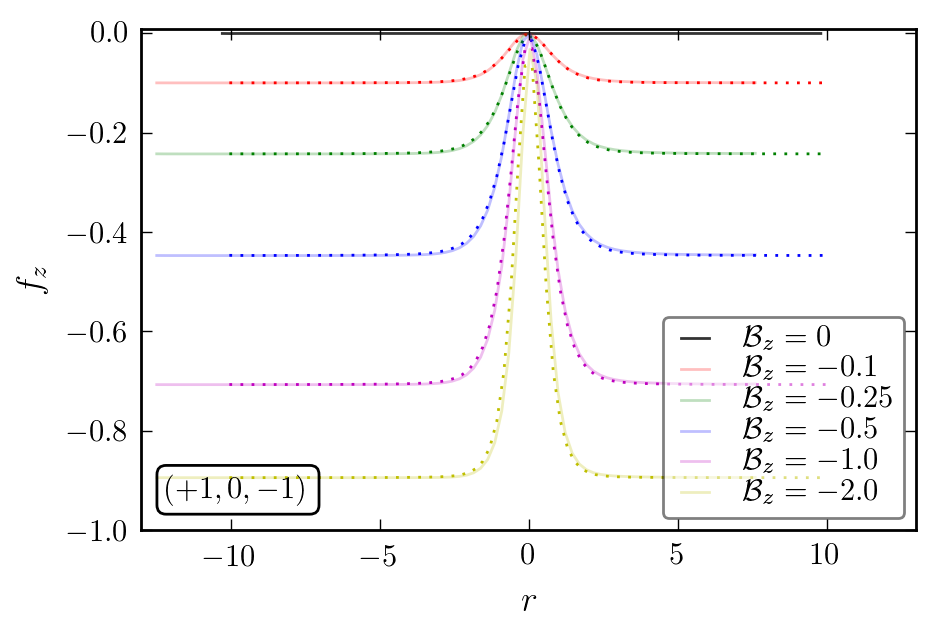}
\caption{Spin texture component $f_z$ as a function of $r$, for different values of $B_z =$ constant, for the vortex structure (+2,+1,0), upper panel, and (+1,0,-1), lower panel. The  dotted lines corresponds to a magnetic field centered at $(x_0 =0,y_0 = 0)$, while the continuous ones to a $(x_0 = 0,y_0 =2)$. This constitutes an indication that $f_z \to 0$ as $r \to \infty$.}  \label{BZ101}
\end{figure}

\begin{figure}[h!]
 \centering
 \includegraphics[width=0.99\linewidth]{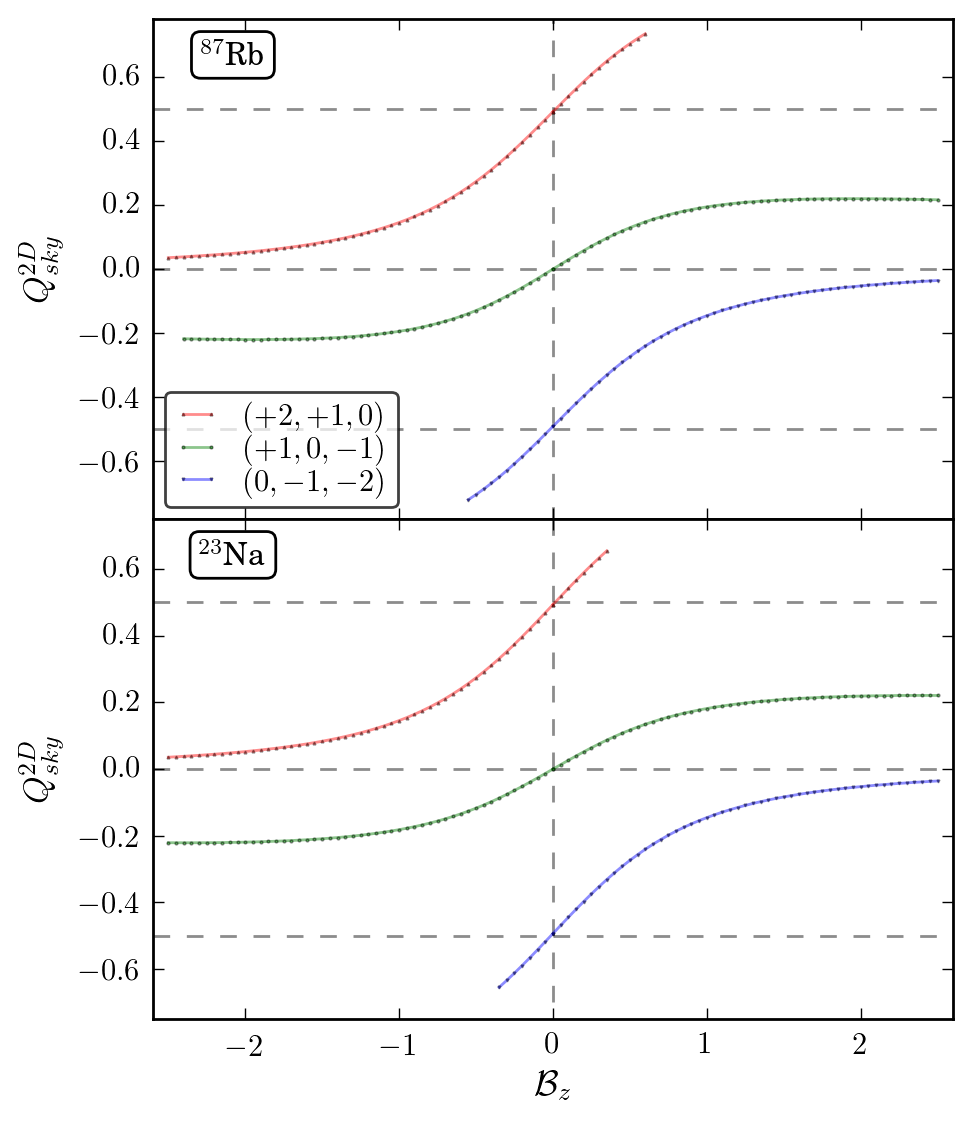}
\caption{(Color online) Skyrmion topological charges $Q_{sky}^{2D}$ as a function of the amplitude ${\cal B}_z$ of the component $B_z = {\cal B}_z r$ of the external magnetic field. Upper panel corresponds to $^{87}$Rb, lower one to $^{23}$Na. 
vortex structures (+2,+1,0), (+1,0,-1), (0,-1-2) are indicated.} \label{topo-arbi}
\end{figure}

We turn our attention now to the case with $B_z = {\cal B}_z r$, which implies that, as $r \to \infty$, the $\vec B$-field can be made to point at any arbitrary direction, depending on the values of ${\cal B}_0$ and ${\cal B}_z$, see Eq.(\ref{boundary}). Figures \ref{BZ101} illustrate typical results of this case. The first refers to vortices (+2,+1,0) and the second one to (+1,0,-1). Again, we have superimposed defects created at $(x_0 = 0, y_0 =0)$ and $(x_0 \ne 0, y_0 \ne 0)$ to check that the asymptotic behavior is correctly inferred: In all cases, we find that as $r \to \infty$, $f_z(r) \not\to 0$, as expected, due to fact that $\vec f$ aligns to the corresponding asymptotic direction of $\vec B$-field. This yields topological charges of arbitrary values. Fig \ref{topo-arbi} summarizes the topological charges obtained for all cases, as the amplitude ${\cal B}_z$ is varied. It shows that the topological charges not only take all values from -1/2 to +1/2, but they can go beyond to higher values, for the corresponding numerically stable cases ${\cal B}_z < 0$ and ${\cal B}_z < 0$) for (+2,+1,0) and (0,-1,-2) respectively. We have verified that for (+2,+1,0) and (0,-1-2), both formulae, Eq.(\ref{magSky}) and (\ref{sky-res}), agree thus supporting our conclusions. The topological nature can again be assured from the standpoint of view that the charge does not depend on the spatial location of the defect and that the charges of several defects add up. 
 
\section{Final Comments}

To the best of our knowledge, there are no previous reports on Skyrmions with arbitrary topological charges, thus, there remains as a task the full elucidation of whether this property can be fully considered as of topological nature. Such a problem is beyond the scope of the present study. Nevertheless, we insist that this result follows very simple from expression (\ref{sky-res}) of the 
$Q_{sky}^{2D}$ Skyrmion topological charge, which tells us that the value can be anything, from $-1/2$ to $-1/2$, depending solely on the boundary values of the spin texture. Moreover, the topological nature of the defects also shows by finding that the values of the charges are independent of the location of the singularity axis and by checking that the charges of multiple defects add up. The simplest explanation is that the the texture becomes paramagnetic away from the location of the defect and, hence, the boundary values can be adjusted by appropriately tailoring the external magnetic field $\vec B$.  The ``arbitrary'' value of this topological charge is also reminiscent of the arbitrariness of Berry phases in spin systems.

An important question left to be addressed is the stability of the states here considered. We have numerically tested this stability criterion as follows. First, we recall that the states are found through a minimization numerical procedure of the energy functional given by Eq. (\ref{totalEnergy}), which is equivalent to solving the time-independent GP equation. Part of this procedure allows to finding the corresponding chemical potential. If the state is stationary, it should not evolve under the propagation of the time-dependent GP. Thus, the states found were evolved for more than 100 time units, approximately 50 milliseconds for the systems here considered, and they remained completely stable for those times. Some of our cases, as mentioned above, have been shown to be stable in in Ref.\cite{PhysRevA.76.023610}. We are thus confident of their stability. 

\begin{acknowledgments}
Acknowledgment is given to grants CONACYT 2555573 (Mexico) and PAPIIT-IN105217 (UNAM). RZZ thanks CONACYT (Mexico) for a graduate scholarship.
\end{acknowledgments}

\bibliographystyle{unsrt}

\bibliography{references}

\begin{thebibliography}{10}

\bibitem{Skyrme127}
T.~H.~R. Skyrme.
\newblock A non-linear field theory.
\newblock {\em Proceedings of the Royal Society of London A: Mathematical,
  Physical and Engineering Sciences}, 260(1300):127--138, 1961.

\bibitem{Brown20101}
G.E. Brown and M.~Rho, editors.
\newblock {\em The multifaceted skyrmion}.
\newblock World Scientific, Singapore, 2010.
\newblock cited By 2.

\bibitem{superconductivity}
Sergey~S. Pershoguba, Sho Nakosai, and Alexander~V. Balatsky.
\newblock Skyrmion-induced bound states in a superconductor.
\newblock {\em Phys. Rev. B}, 94:064513, Aug 2016.

\bibitem{Nagaosa2013899}
N.~Nagaosa and Y.~Tokura.
\newblock Topological properties and dynamics of magnetic skyrmions.
\newblock {\em Nature Nanotechnology}, 8(12):899--911, 2013.
\newblock cited By 322.

\bibitem{Mansoor2014}
Mansoor B.~A. Jalil and Seng~Ghee Tan.
\newblock Robustness of topological hall effect of nontrivial spin textures.
\newblock {\em Scientific Reports}, 4:5123 EP --, 05 2014.

\bibitem{liquidcrystals}
A.~O. Leonov, I.~E. Dragunov, U.~K. R\"o\ss{}ler, and A.~N. Bogdanov.
\newblock Theory of skyrmion states in liquid crystals.
\newblock {\em Phys. Rev. E}, 90:042502, Oct 2014.

\bibitem{Leonhardt2000}
U.~Leonhardt and G.~E. Volovik.
\newblock How to create an alice string (half-quantum vortex) in a vector
  bose-einstein condensate.
\newblock {\em Journal of Experimental and Theoretical Physics Letters},
  72(2):46--48, 2000.

\bibitem{PhysRevA.62.013602}
Karl-Peter Marzlin, Weiping Zhang, and Barry~C. Sanders.
\newblock Creation of skyrmions in a spinor bose-einstein condensate.
\newblock {\em Phys. Rev. A}, 62:013602, Jun 2000.

\bibitem{PhysRevLett.88.090404}
J.-P. Martikainen, A.~Collin, and K.-A. Suominen.
\newblock Creation of a monopole in a spinor condensate.
\newblock {\em Phys. Rev. Lett.}, 88:090404, Feb 2002.

\bibitem{skyrmionOnHalfSpinor}
U.~Al~Khawaja and H.~Stoof.
\newblock Skyrmions in a ferromagnetic bose - einstein condensate.
\newblock {\em Nature}, 411(6840):918--920, 2001.
\newblock cited By 155.

\bibitem{PhysRevLett.88.080401}
Richard~A. Battye, N.~R. Cooper, and Paul~M. Sutcliffe.
\newblock Stable skyrmions in two-component bose-einstein condensates.
\newblock {\em Phys. Rev. Lett.}, 88:080401, Feb 2002.

\bibitem{PhysRevA.93.033633}
Justin Lovegrove, Magnus~O. Borgh, and Janne Ruostekoski.
\newblock Stability and internal structure of vortices in spin-1 bose-einstein
  condensates with conserved magnetization.
\newblock {\em Phys. Rev. A}, 93:033633, Mar 2016.

\bibitem{PhysRevLett.109.015301}
Takuto Kawakami, Takeshi Mizushima, Muneto Nitta, and Kazushige Machida.
\newblock Stable skyrmions in $su(2)$ gauged bose-einstein condensates.
\newblock {\em Phys. Rev. Lett.}, 109:015301, Jul 2012.

\bibitem{PhysRevLett.100.180403}
Yuki Kawaguchi, Muneto Nitta, and Masahito Ueda.
\newblock Knots in a spinor bose-einstein condensate.
\newblock {\em Phys. Rev. Lett.}, 100:180403, May 2008.

\bibitem{Liu20133300}
Yong-Kai Liu, Cong Zhang, and Shi-Jie Yang.
\newblock 3d skyrmion and knot in two-component bose--einstein condensates.
\newblock {\em Physics Letters A}, 377(45--48):3300 -- 3303, 2013.

\bibitem{BorghPRL2016}
Magnus~O. Borgh, Muneto Nitta, and Janne Ruostekoski.
\newblock Stable core symmetries and confined textures for a vortex line in a
  spinor bose-einstein condensate.
\newblock {\em Phys. Rev. Lett.}, 116:085301, Feb 2016.

\bibitem{PhysRevLett.103.250401}
L.~S. Leslie, A.~Hansen, K.~C. Wright, B.~M. Deutsch, and N.~P. Bigelow.
\newblock Creation and detection of skyrmions in a bose-einstein condensate.
\newblock {\em Phys. Rev. Lett.}, 103:250401, Dec 2009.

\bibitem{PhysRevLett.108.035301}
Jae-yoon Choi, Woo~Jin Kwon, and Yong-il Shin.
\newblock Observation of topologically stable 2d skyrmions in an
  antiferromagnetic spinor bose-einstein condensate.
\newblock {\em Phys. Rev. Lett.}, 108:035301, Jan 2012.

\bibitem{Hall2016}
D.S. Hall, M.W. Ray, K.~Tiurev, E.~Ruokokoski, A.H. Gheorghe, and
  M.~M{\"o}tt{\"o}nen.
\newblock Tying quantum knots.
\newblock {\em Nature Physics}, 12(5), 2016.
\newblock cited By 2.

\bibitem{Ray2014657}
M.W. Ray, E.~Ruokokoski, S.~Kandel, M.~M{\"o}tt{\"o}nen, and D.S. Hall.
\newblock Observation of dirac monopoles in a synthetic magnetic field.
\newblock {\em Nature}, 505(7485):657--660, 2014.
\newblock cited By 63.

\bibitem{LeanhardtPRL2003}
A.~E. Leanhardt, Y.~Shin, D.~Kielpinski, D.~E. Pritchard, and W.~Ketterle.
\newblock Coreless vortex formation in a spinor bose-einstein condensate.
\newblock {\em Phys. Rev. Lett.}, 90:140403, Apr 2003.

\bibitem{ChoiNJP2012}
Jae yoon Choi, Woo~Jin Kwon, Moonjoo Lee, Hyunseok Jeong, Kyungwon An, and Yong
  il~Shin.
\newblock Imprinting skyrmion spin textures in spinor bose--einstein
  condensates.
\newblock {\em New Journal of Physics}, 14(5):053013, 2012.

\bibitem{XuPRA2012}
Xiao-Qiang Xu and Jung~Hoon Han.
\newblock Skyrmion dynamics and disintegration in a spin-1 bose-einstein
  condensate.
\newblock {\em Phys. Rev. A}, 86:063619, Dec 2012.

\bibitem{HuangPRA2013}
Chao-Chun Huang and S.-K. Yip.
\newblock Dynamics and complex structure of two-dimensional skyrmions in
  antiferromagnetic spin-1 bose-einstein condensates.
\newblock {\em Phys. Rev. A}, 88:013628, Jul 2013.

\bibitem{RevModPhys.51.591}
N.~D. Mermin.
\newblock The topological theory of defects in ordered media.
\newblock {\em Rev. Mod. Phys.}, 51:591--648, Jul 1979.

\bibitem{topoBook}
Mikio Nakahara.
\newblock {\em Geometry, topology, and physics}.
\newblock Graduate student series in physics. Institute of Physics Publishing,
  Bristol, Philadelphia, 2003.

\bibitem{uedaAspectsSBEC}
Masahito Ueda.
\newblock Topological aspects in spinor bose--einstein condensates.
\newblock {\em Reports on Progress in Physics}, 77(12):122401, 2014.

\bibitem{mhVortex}
T.~Mizushima, K.~Machida, and T.~Kita.
\newblock Mermin-ho vortex in ferromagnetic spinor bose-einstein condensates.
\newblock {\em Phys. Rev. Lett.}, 89:030401, Jun 2002.

\bibitem{vorticesTopo}
KENICHI KASAMATSU, MAKOTO TSUBOTA, and MASAHITO UEDA.
\newblock Vortices in multicomponent bose--einstein condensates.
\newblock {\em International Journal of Modern Physics B}, 19(11):1835--1904,
  2005.

\bibitem{baby}
T.~Gisiger and M.B. Paranjape.
\newblock Baby skyrmion strings.
\newblock {\em Physics Letters B}, 384(1):207 -- 212, 1996.

\bibitem{HuPRA2015}
Yu-Xin Hu, Christian Miniatura, and Beno\^{\i}t Gr\'emaud.
\newblock Half-skyrmion and vortex-antivortex pairs in spinor condensates.
\newblock {\em Phys. Rev. A}, 92:033615, Sep 2015.

\bibitem{SuPRA2012}
S.-W. Su, I.-K. Liu, Y.-C. Tsai, W.~M. Liu, and S.-C. Gou.
\newblock Crystallized half-skyrmions and inverted half-skyrmions in the
  condensation of spin-1 bose gases with spin-orbit coupling.
\newblock {\em Phys. Rev. A}, 86:023601, Aug 2012.

\bibitem{RuostekoskiPRL2001}
J.~Ruostekoski and J.~R. Anglin.
\newblock Creating vortex rings and three-dimensional skyrmions in
  bose-einstein condensates.
\newblock {\em Phys. Rev. Lett.}, 86:3934--3937, Apr 2001.

\bibitem{Zeng2009854}
Rong Zeng and Yanzhi Zhang.
\newblock Efficiently computing vortex lattices in rapid rotating
  bose--einstein condensates.
\newblock {\em Computer Physics Communications}, 180(6):854 -- 860, 2009.

\bibitem{RevModPhys.85.1191}
Dan~M. Stamper-Kurn and Masahito Ueda.
\newblock Spinor bose gases: Symmetries, magnetism, and quantum dynamics.
\newblock {\em Rev. Mod. Phys.}, 85:1191--1244, Jul 2013.

\bibitem{hoSpinor}
Tin-Lun Ho.
\newblock Spinor bose condensates in optical traps.
\newblock {\em Phys. Rev. Lett.}, 81:742--745, Jul 1998.

\bibitem{nosostrosOnDemand}
R.~Zamora-Zamora, M.~Lozada-Hidalgo, S.~F. Caballero-Ben\'{\i}tez, and
  V.~Romero-Roch\'{\i}n.
\newblock Vortices on demand in multicomponent bose-einstein condensates.
\newblock {\em Phys. Rev. A}, 86:053624, Nov 2012.

\bibitem{PhysRevA.76.023610}
V.~Pietil\"a, M.~M\"ott\"onen, and S.~M.~M. Virtanen.
\newblock Stability of coreless vortices in ferromagnetic spinor bose-einstein
  condensates.
\newblock {\em Phys. Rev. A}, 76:023610, Aug 2007.

\bibitem{PhysRevA.61.063610}
Tomoya Isoshima, Mikio Nakahara, Tetsuo Ohmi, and Kazushige Machida.
\newblock Creation of a persistent current and vortex in a bose-einstein
  condensate of alkali-metal atoms.
\newblock {\em Phys. Rev. A}, 61:063610, May 2000.

\bibitem{calculation}
To perform our calculations we use state of the art GPU programming via the
  PyCUDA library in a GPU with 2304 CUDA cores. We do calculations with double
  and single precision on $128^3$ and $256^3$ mesh points. We point out that
  the structure of the stationary states can be numerically ensured with
  $128^3$ within single precision calculations.

\end{thebibliography}

\end{document}